\documentclass[showpacs,longbibliography]{svproc}
\usepackage[utf8]{inputenc}

\usepackage{graphicx}
\usepackage{multicol}
\usepackage{footmisc}
\usepackage[english]{babel}
\usepackage{float}
\usepackage{amsmath,amssymb,enumerate,bm}
\usepackage{url,bbding}

\title{Acoustic resonances in a confined set of macroscopic disks}
\author{Juan F. González-Saavedra$^\text{\Envelope}$, \'Alvaro Rodríguez-Rivas, Miguel A. López-Castaño
  and Francisco Vega Reyes}
\authorrunning{González-Saavedra \textit{et al.}}
\institute{Departamento de Física and Instituto de Computación Científica Avanzada (ICCAEx), Universidad de Extremadura, 06071 Badajoz, Spain
\\ \small{e-mail: \email{juanfrangs@unex.es}}}
\date{\today}

\begin{document}

\maketitle

%\section{Abstract}
\begin{abstract}
  We study in this work a system of granular disks enclosed in a rectangular region. Granular disks are arranged, at high particle density,  in a hexagonal lattice. Specifically, we are interested in the conditions for mechanical signal transmission. Two different potentials are considered. Our results are obtained by means of soft disks computer simulations. We analytically determine the eigenfrequencies spectra of the system from the corresponding Hessian matrix.  %As we will show, the efficiency. 
  Alternatively, the eigenfrequencies can also be obtained from the time evolution
  of molecular dynamics simulations. Previous works have shown that mechanical signal transmission in granular matter can be used to develop acoustic switches. Our results reveal that the working range of the granular switch changes dramatically for different interaction models. In particular, strong anisotropies in signal transmission may appear for certain interparticle forces. 
\end{abstract}

\section{Introduction}

Signal transmission devices are ubiquitous in mechanical engineering
applications \cite{LGYSL20,DCRPT19}. They are used as vibration insulators
\cite{GSHMD13}, acoustic cloaks \cite{ZPC14}, sound propagation \cite{C14}, signal
rectification \cite{BTDK11}, etc. In
this context, highly dense granular systems can be used to perform these tasks \cite{WCBSO19}. In
particular, granular crystals have proven to be efficient for clean signal
transmission \cite{BTD11}. In fact, some recent works show that they can work as switches \cite{WCBSO19},
rectifiers \cite {BTD11} and a range of logic elements \cite{MFNFY11}.

Hence, granular crystals can be used to modify and transmit mechanical energy signals
\cite{AB13}. These applications are analogous to the electrical switches that can
control electromagnetic waves propagation \cite{BB48}. Control of acoustic and thermal
information using this type of switches can lead to applications in fields such as
thermoelectric systems, communication or ultrasonic imaging \cite{LAYKD14}. 

As it is known, dense granular systems of identically shaped particles can develop crystalline structures \cite{Melby2005}. Usually, the critical density for the appearance of ordering  depends strongly on the values of the other relevant
physical magnitudes: granular temperature \cite{PMEU04}, degree of inelasticity in collisions \cite{VU08}, the
specifics of the physical processes involved in particle collisions \cite{OU98}, the type of
interaction with the boundaries \cite{LVU09}, the kind of energy input necessary to thermalize the
granular particles \cite{LGRV19}, etc. Experimental and computational work has put in evidence
these effects. 

We study in
this work a system consisting of macroscopic disks by means of molecular dynamics
simulations. In order to assess the influence of interparticle
collisions,  we consider two interaction potentials: a potential that is quadratic in
the distance between particles \cite{WCBSO19}, and a Lennard-Jones-type potential \cite{R04}

The paper is organized as follows. The second section is devoted to the system
description, including the initial configuration used in this work, together with the
computational and theoretical methods used to obtain the results. Finally, Section 3 presents
the results and a discussion, together with a description of possible future extensions
of our results.

%EQUATIONS, POTENTIALS

\section{Description of the system and computational methods}

\begin{figure}[ht]
\centering
\includegraphics[width=.350 \columnwidth]{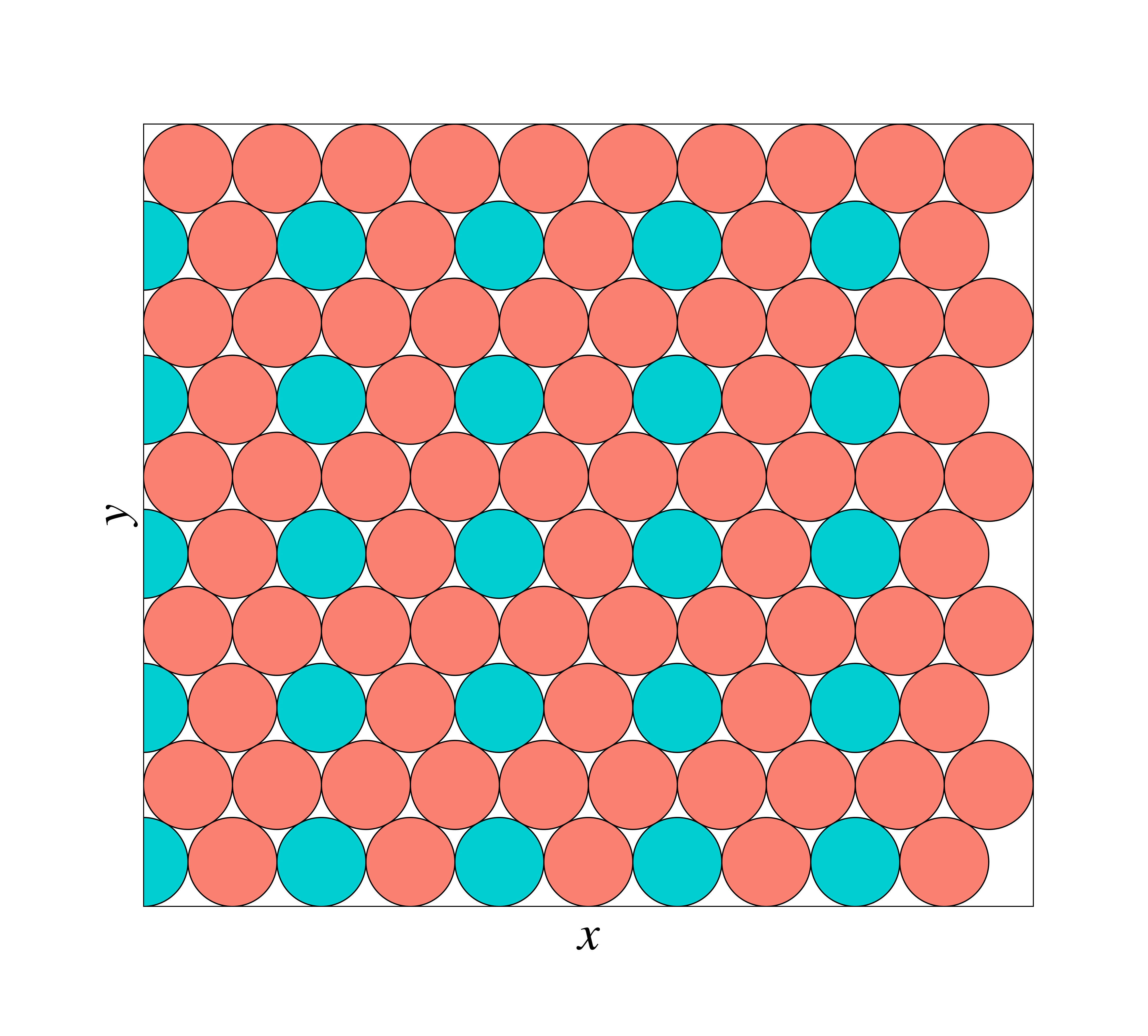}
\caption{Initial setup of the light (blue) and heavy (red) disks, with $10\times10$ particle layers. $x$ stands for the horizontal axis and
$y$ for the vertical axis.} \label{Setup}
\end{figure}

Let us imagine a rectangular region defined by two parallel rigid walls and two parallel periodic boundaries, as sketched in Fig. \ref{Setup}. There is no gravitational field acting on the system. Rigid walls-disks interactions are modelled with the same potential as for disk-disk interactions. More specifically, rigid walls are modelled by assuming the existence of a particle of the same size and mass behind the wall, when the wall is within the interaction range of a real particle. The $x$ position of this wall particle coincides at all times with the $x$ position of the corresponding real particle. On the contrary, the $y$ position of the wall particle center is always located at a particle radius away from the wall surface.

Two different types of interaction are used. The first one is parabolic-like, with the form 

	\begin{equation}\label{eqn:wu}
	U(r_{ij})= \left\{ \begin{array}{cc}
	\frac{\epsilon}{2}\left(1-\frac{r_{ij}}{\sigma}\right)^{2} &    \quad\mathrm{if} \quad r_{ij} \leq \sigma \\
	\\ 0 &\quad\mathrm{if}  \quad r_{ij} > \sigma \\

	\end{array}
	\right. 
	\end{equation}
	whereas the second potential is a Lennard-Jones type potential \cite{R04}

		\begin{equation}\label{eqn:lj}
	U(r_{ij})= \left\{ \begin{array}{cc}
	4\epsilon \left[ \left(\frac{\sigma}{r_{ij}}\right)^{12} -\left(\frac{\sigma}{r_{ij}}\right)^{6} \right]+\epsilon& \quad \mathrm{if}   \quad r_{ij} \leq r_{c} \\
	\\ 0 & \quad \mathrm{if} \quad r_{ij} > r_{c} \, ,\\
	
	\end{array}
	\right.
	\end{equation}
\smallskip
where $\epsilon$ sets the energy scale, $\sigma$ is the diameter of the disks, $r_{ij}$ is the distance between centers of disks $i$ and $j$ and $r_c$ is the energy cut distance, set in our simulations to $2^{1/6}\sigma$.

In order to obtain the eigenfrequencies of the system, we computed the eigenvalues $\omega_{k}^{2}$ of the dynamical matrix $\mathcal{D}$, whose elements are \cite{WCBSO19,TWLB02}

	\begin{equation} \label{eqn:mwdm}
		\mathcal{D}_{ij} = \mathcal{M}^{-1}_{ik}\mathcal{H}_{kj}
		\end{equation}
		with
		\begin{equation} 
		\mathcal{H}_{ij}=\frac{\partial^{2} U}{\partial x_{i} \partial x_{j}}, \qquad	\mathcal{M}_{ij} = m_{L,H}\delta_{ij}
		\end{equation} where $\mathcal{M}_{ij}$, $\mathcal{H}_{ij}$ are the corresponding Hessian matrix \cite{TWLB02} and the particle mass diagonal matrices, respectively, and $m_{L}, m_{H}$ are the masses of the particles. The eigenvalues were obtained by: a) considering an initial equilibrium state of the system; b) from the average positions of the particles for a simulation run.

%It is remarkable the existence of a gap. This is precisely the fact that allows its behavior as a switch. It is possible to choose the operation regime,determined by the frequency, by modifying some variables such as the pressure.

In the simulations, a 2D system of particles (disks with equal diameter $\sigma$) inside the rectangular box is considered. The dimension of the box is such that an integer number of horizontal layers can be perfectly accommodated into a hexagonal lattice, as it can be seen in Fig. \ref{Setup}.
The total number of particles is 100, each of them with a diameter $\sigma$, which we take as length unit. %Moreover, the particles are soft, which means they can overlap other particles and the walls.%
We consider particles with two different masses, the mass ratio being $m_{H}/m_{L}=10$. The particle distribution is such that we have 25 light disks (with mass $m_L$) and 75 heavy disks (with mass $m_H$). The initial configuration is the mechanically stable packing of the disks. 

%\begin{figure}[ht]
%\includegraphics[width=.50 \columnwidth]{./figs/evparabolaT0.png}
%\includegraphics[width=.50 \columnwidth]{./figs/evljT0.png}
%\caption{Eigenfrequencies of the system when the parabolic (left panel) and Lennard-Jones %(right panel) potentials are considered for $K=10^{-4}$ (blue) and $K=10^{-2}$ (red).} %\label{eigenfreq_par}
%\end{figure}

%%#\begin{figure}[ht]
%%#\includegraphics[width=.85 \columnwidth]{./figs/evljT0.png}
%%\caption{Eigenfrequencies of the system when the Lennard-Jones potential is considered.} %%\label{eigenfreq_lj}
%%\end{figure}

Initial particle velocities are drawn from a Maxwell-Boltzmann distribution whose standard deviation is determined by the system initial temperature. In our case, this temperature is a control parameter in the simulations.
 The evolution of the system is obtained from molecular dynamics simulations using leap-frog method \cite{R04}. The pressure can also be modified by approaching the rigid walls to each other.

%\begin{itemize}
%\item $100$ particles with the same diameter Two different masses with ratio $10$: light (blue) and heavy (red)

%\item Initial velocity of the disks given by an input temperature value
%\item Disk-disk and disk-wall interactions according to potentials $1$ and $2$
%\item Molecular dynamics simulation using leap-frog method
%  $20Millions$ of iterations
%\end{itemize}

\section{Results and discussion}
%\section{VELOCITY AUTOCORRELATION FUNCTION}

\begin{figure}[ht]
\centering
\includegraphics[width=.45\columnwidth]{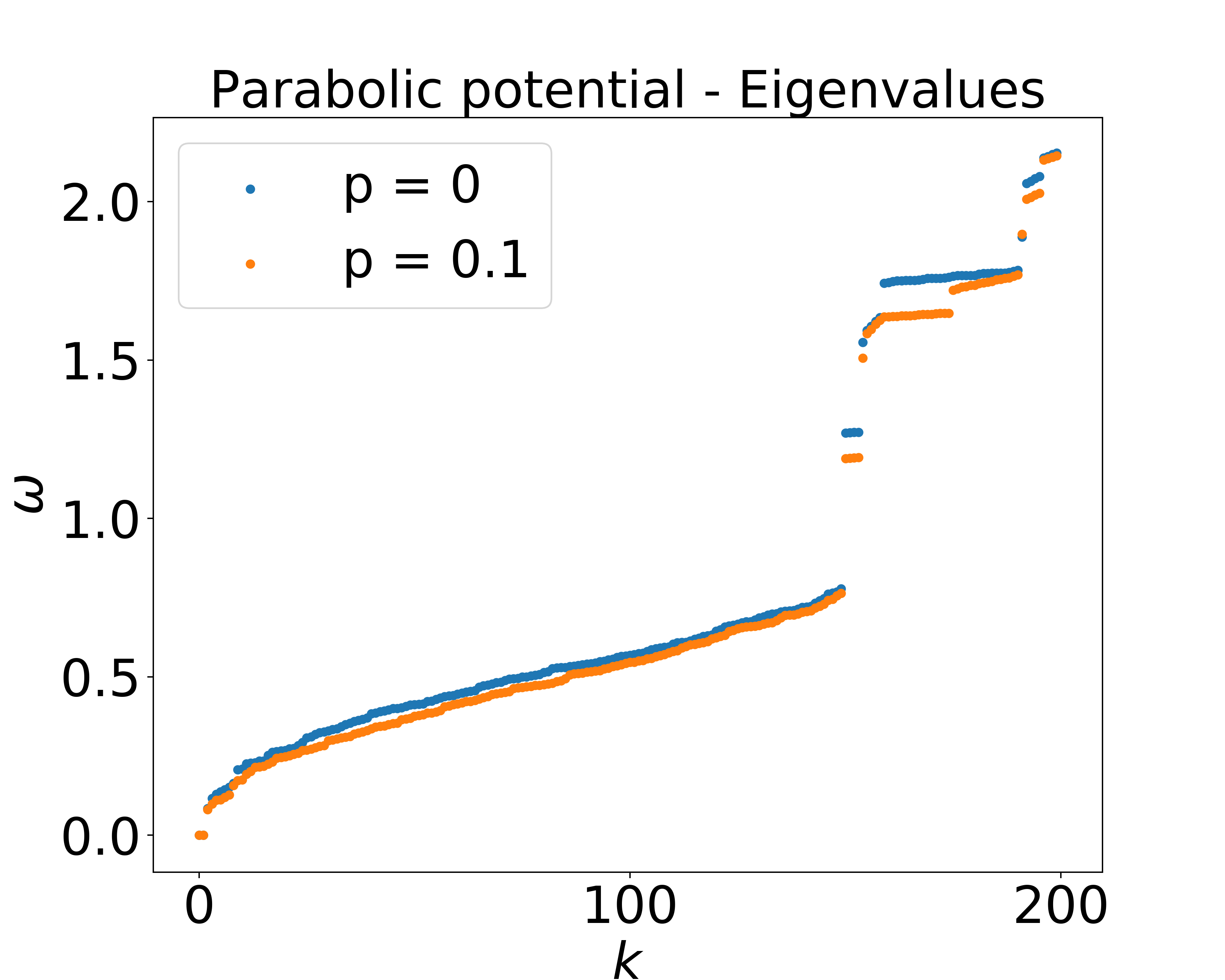}\quad
\includegraphics[width=.45\columnwidth]{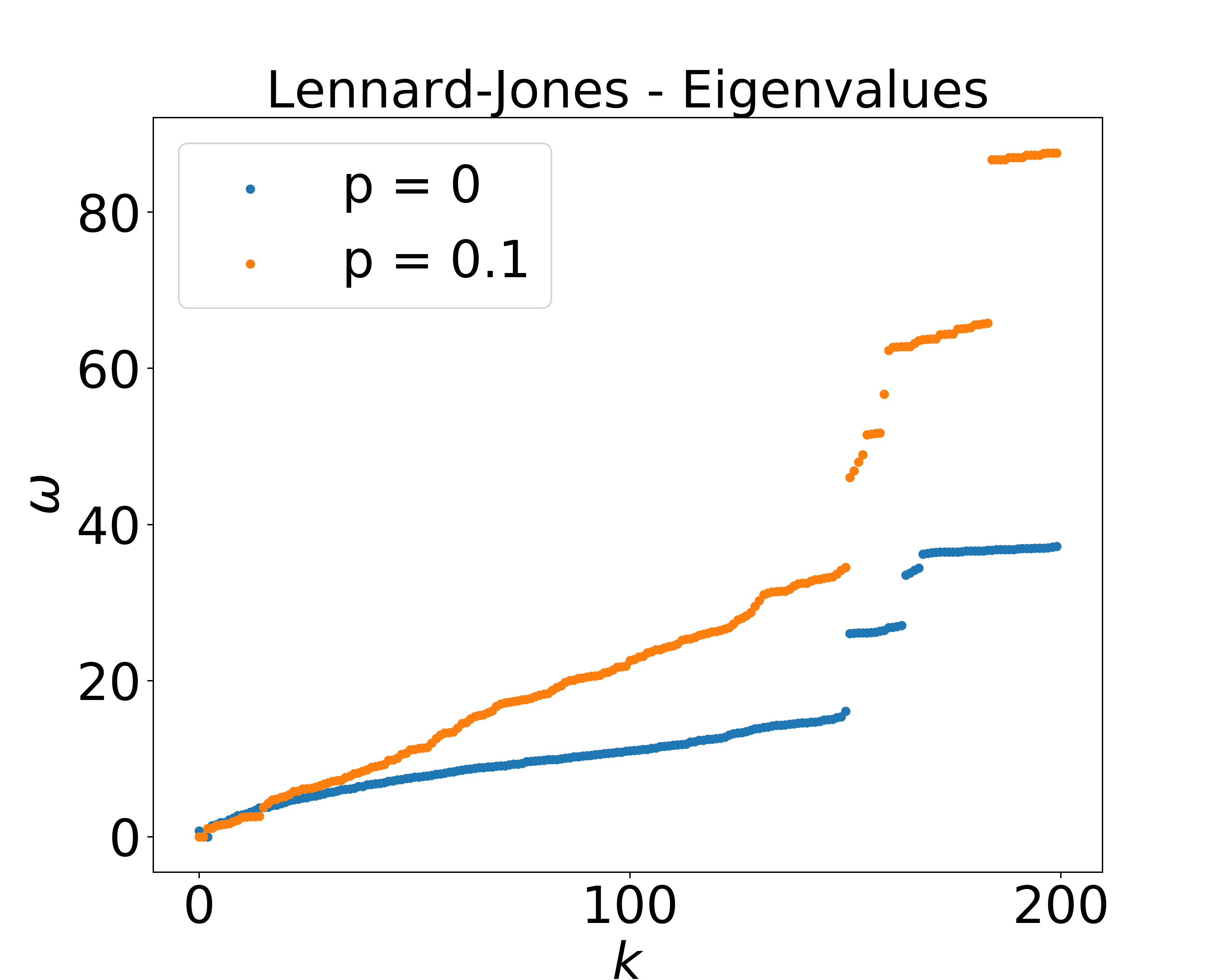}
\caption{Eigenfrequencies of the system when the parabolic (left panel) and Lennard-Jones (right panel) potentials are considered. The curves represent a system with the particles in their equilibrium positions (blue) and with a pressure $p=10^{-1}$ (red).} \label{eigenfreq_par}
\end{figure}

In Fig. \ref{eigenfreq_par} we can see the corresponding eigenvalues for both parabolic \eqref{eqn:wu} and Lennard-Jones \eqref{eqn:lj} potentials. The equilibrium positions are considered as $p=0$ configurations. This state can be modified by approaching the top and bottom walls. In this case, the disks will overlap, and, as a consequence, repulsion forces between them appear. By means of this procedure, we can slightly increase the pressure to $p=0.1$. As we can see, the eigenvalues display significant discontinuities. This produces frequency gaps which can be tuned by modifying the pressure of the system. It is this behavior which allows for the use as a switch. Put in other words, a mechanical signal with a given frequency  $\omega$ will be transmitted only if it does not lie within a forbidden frequency gap (for a given value of the system pressure $p$).

\begin{figure}[ht]
\centering
\includegraphics[width=.45 \columnwidth]{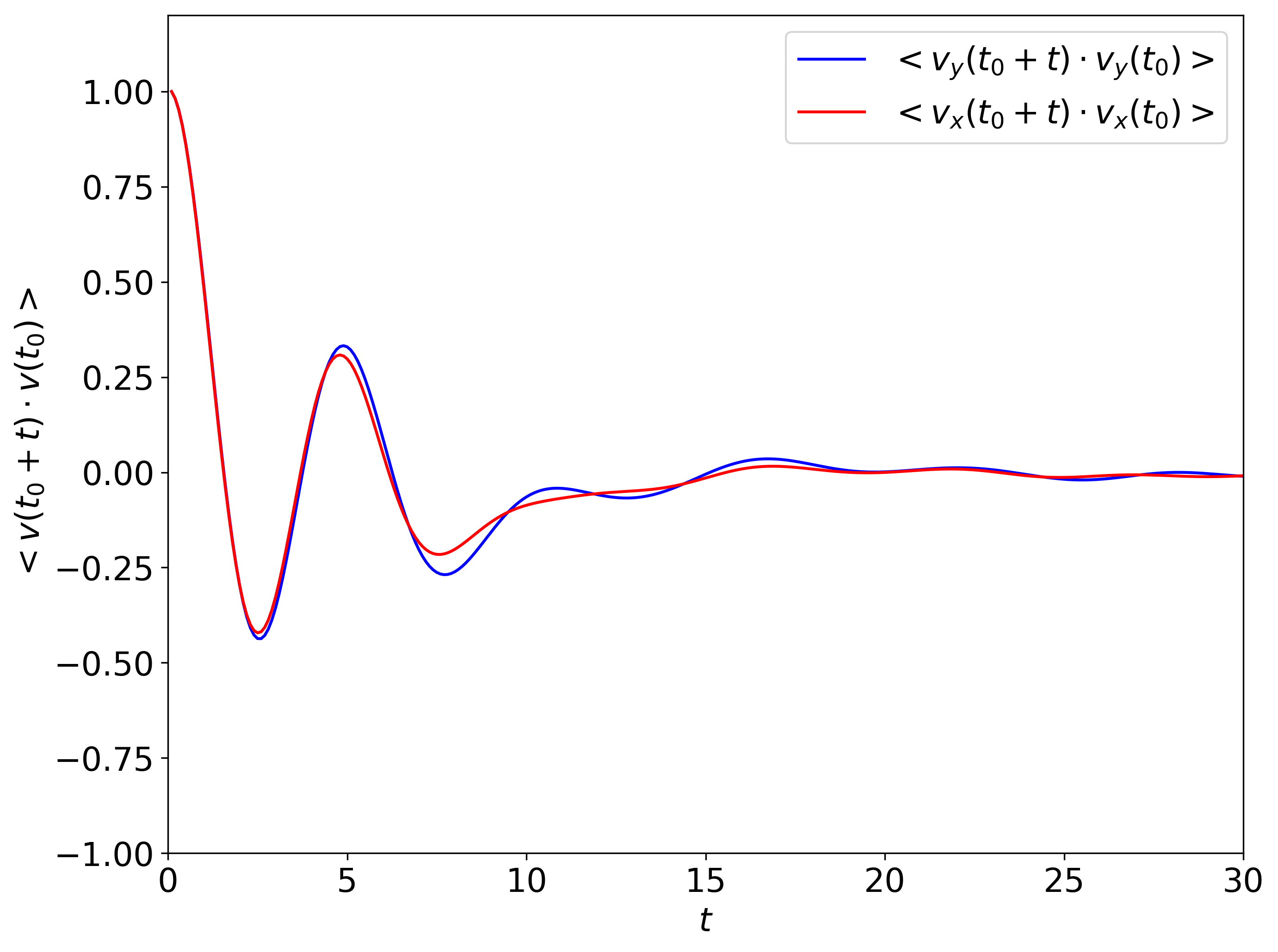}\quad
\includegraphics[width=.47 \columnwidth]{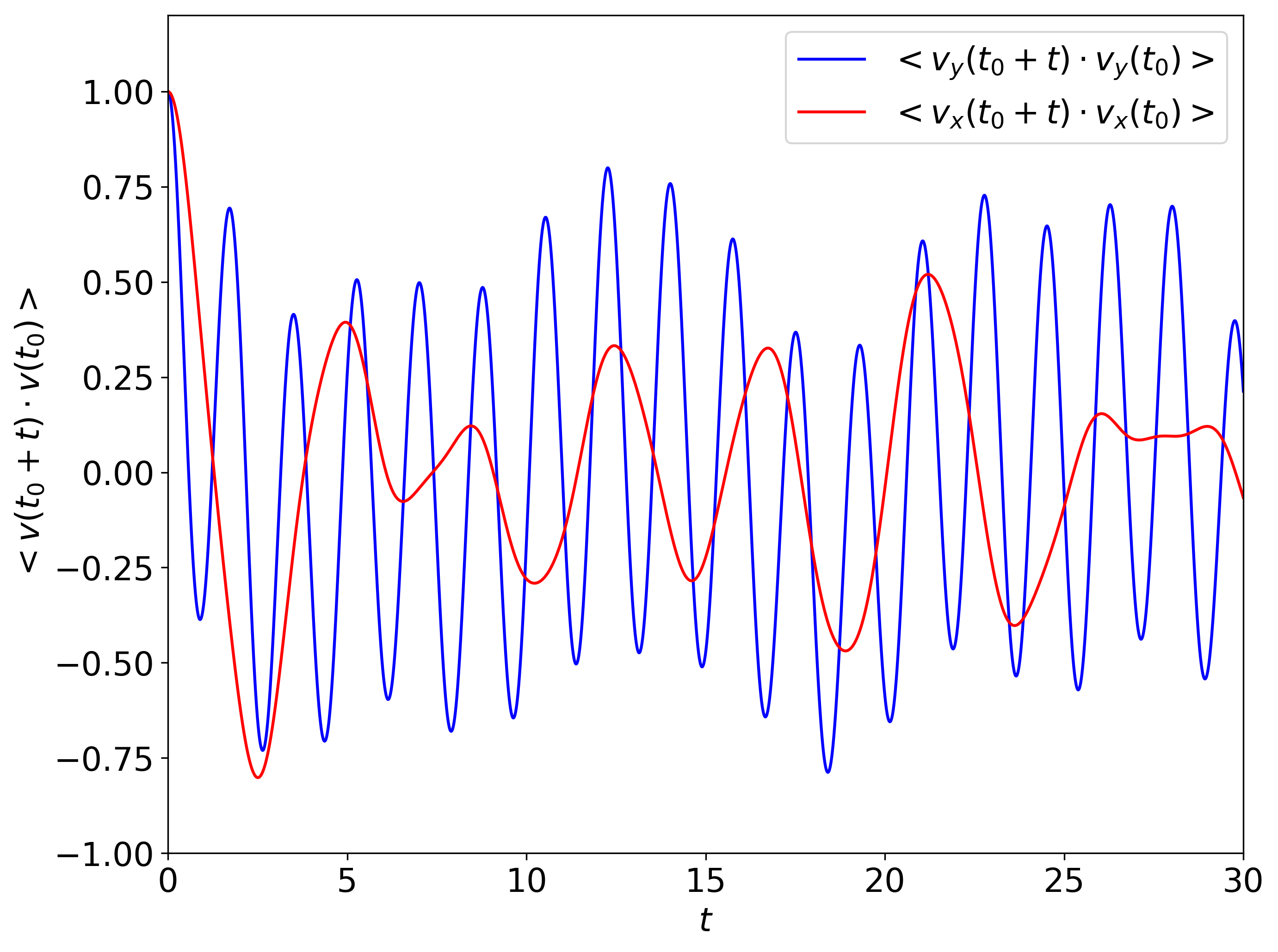}
\caption{Velocity autocorrelation function in the $x$ and $y$ directions for the parabolic (left panel) and the Lennard-Jones potentials (right panel) at temperature $T=10^{-4}$.} \label{Velocity}
\end{figure}

As we can see in Fig. \ref{Velocity} the normalized velocity autocorrelation function ($\langle \mathbf{v}(t_0+t)\cdot \mathbf{v} (t_0)\rangle/\langle \mathbf{v}(t_0)\cdot \mathbf{v} (t_0)\rangle$) has finite decay length and is isotropic for the parabolic potential whereas
for the Lennard-Jones potential does not decay and there are significant differences between the $x$ and $y$ directions. The peculiar behavior in the case of the Lennard-Jones may be due to its steep repulsive term, as compared to the parabolic potential.

\begin{figure}[ht]
\centering
\includegraphics[width=.45 \columnwidth]{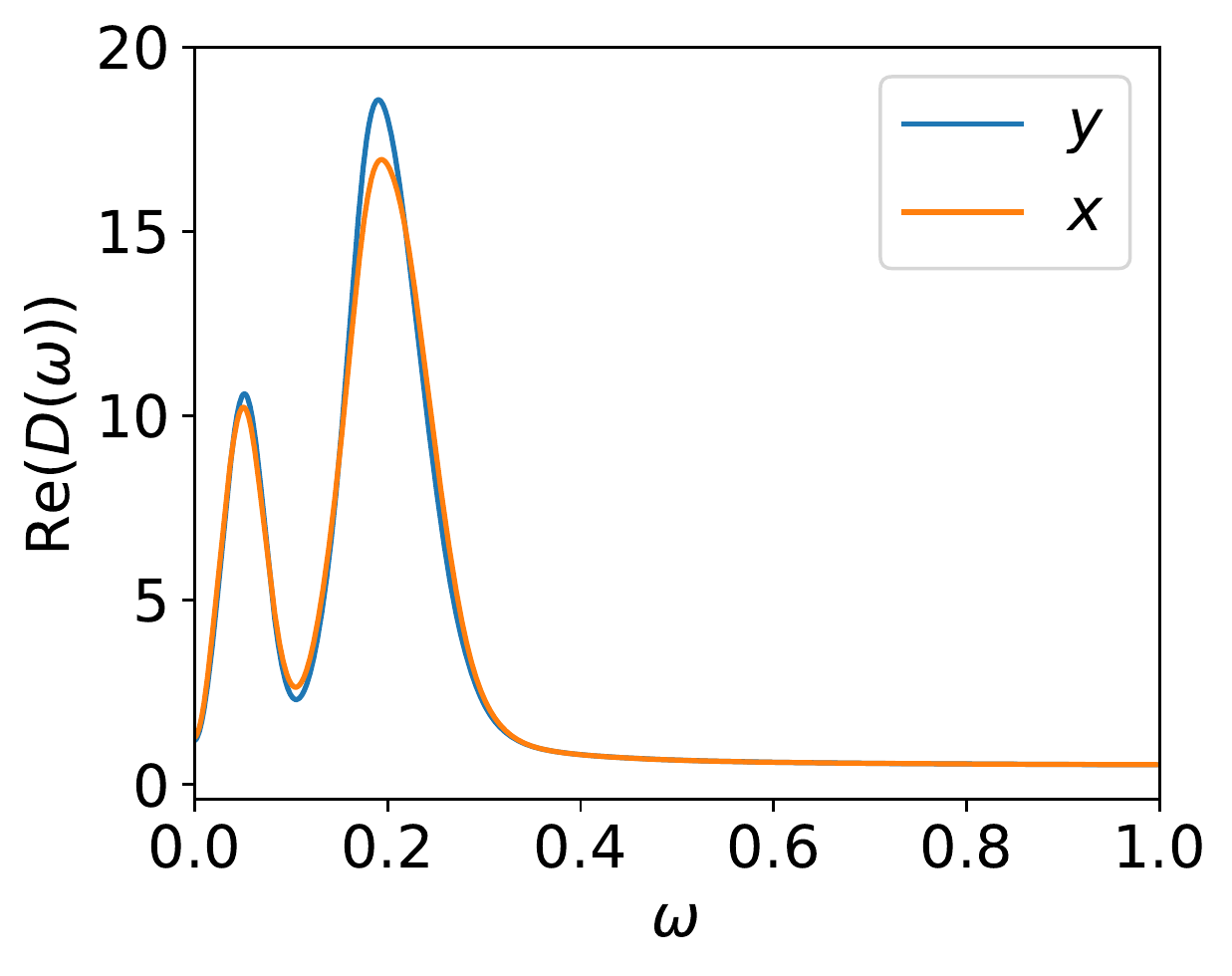}\quad
\includegraphics[width=.45 \columnwidth]{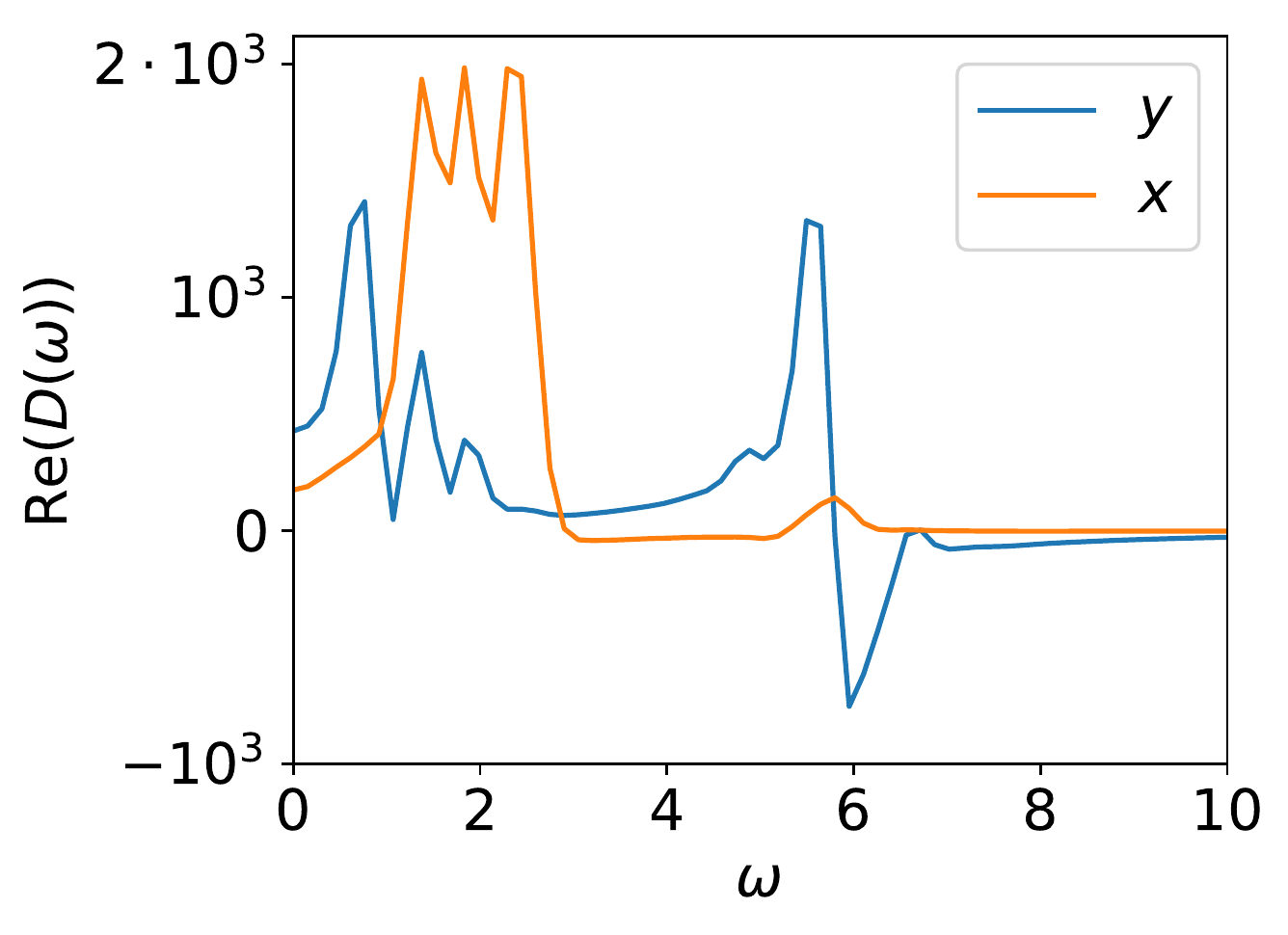}
\caption{Power spectra distribution (Fourier transform of the velocity autocorrelations) in the $x$ and $y$ directions for the parabolic (left panel) and the Lennard-Jones potentials (right panel) at temperature $T=10^{-4}$.} \label{fourier}
\end{figure}

A way to look at the behavior of an eventual granular switch is to calculate the power spectra density
\begin{equation}
D(\omega)=\int_{0}^{\infty}{dt\frac{\langle \mathbf{v}(t_0+t)\cdot \mathbf{v} (t_0)\rangle}{\langle \mathbf{v}(t_0)\cdot \mathbf{v} (t_0)\rangle}e^{i\omega t}}
\end{equation}

\begin{figure}[ht]
\centering
\includegraphics[height= 4cm ]{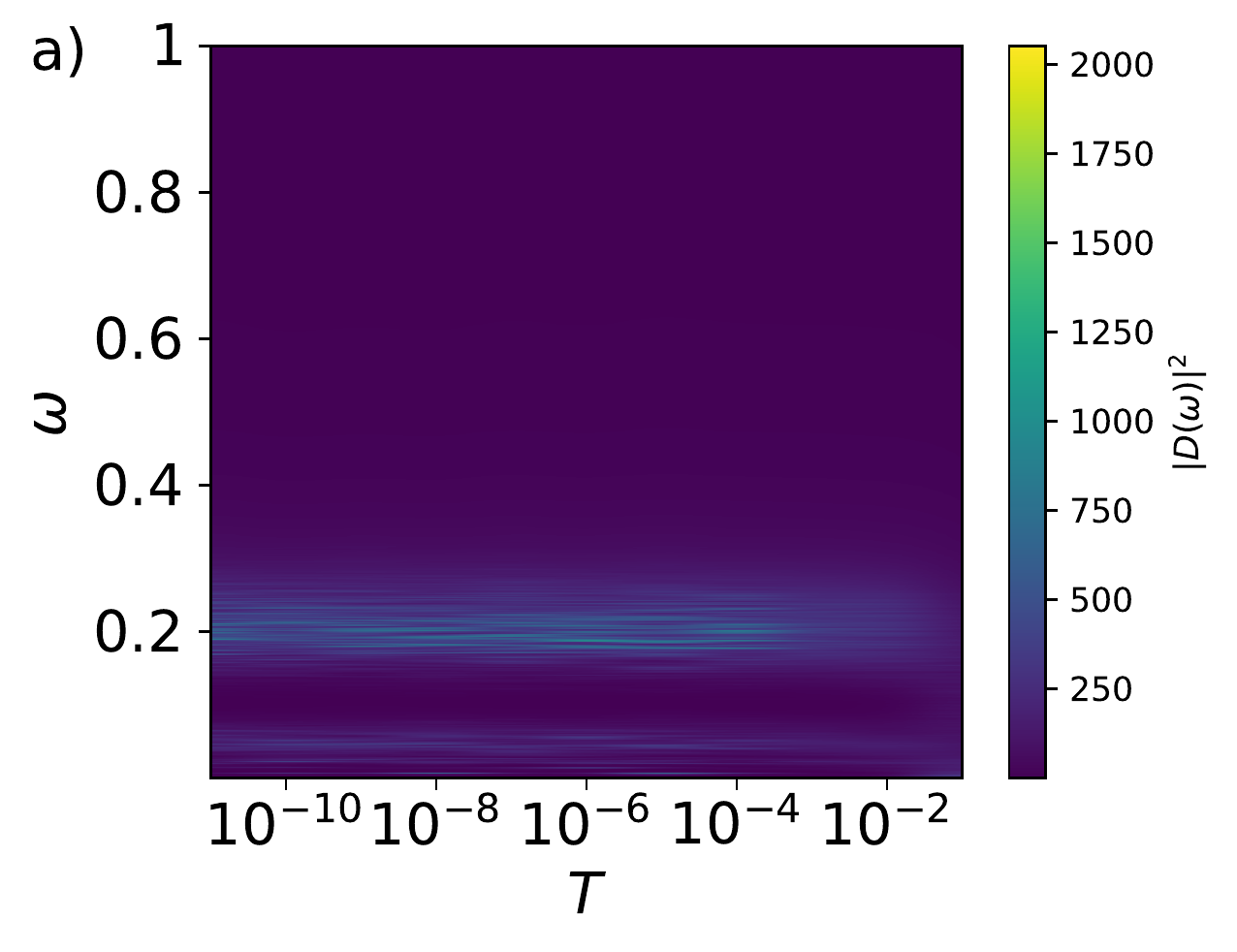}
\includegraphics[height= 4cm ]{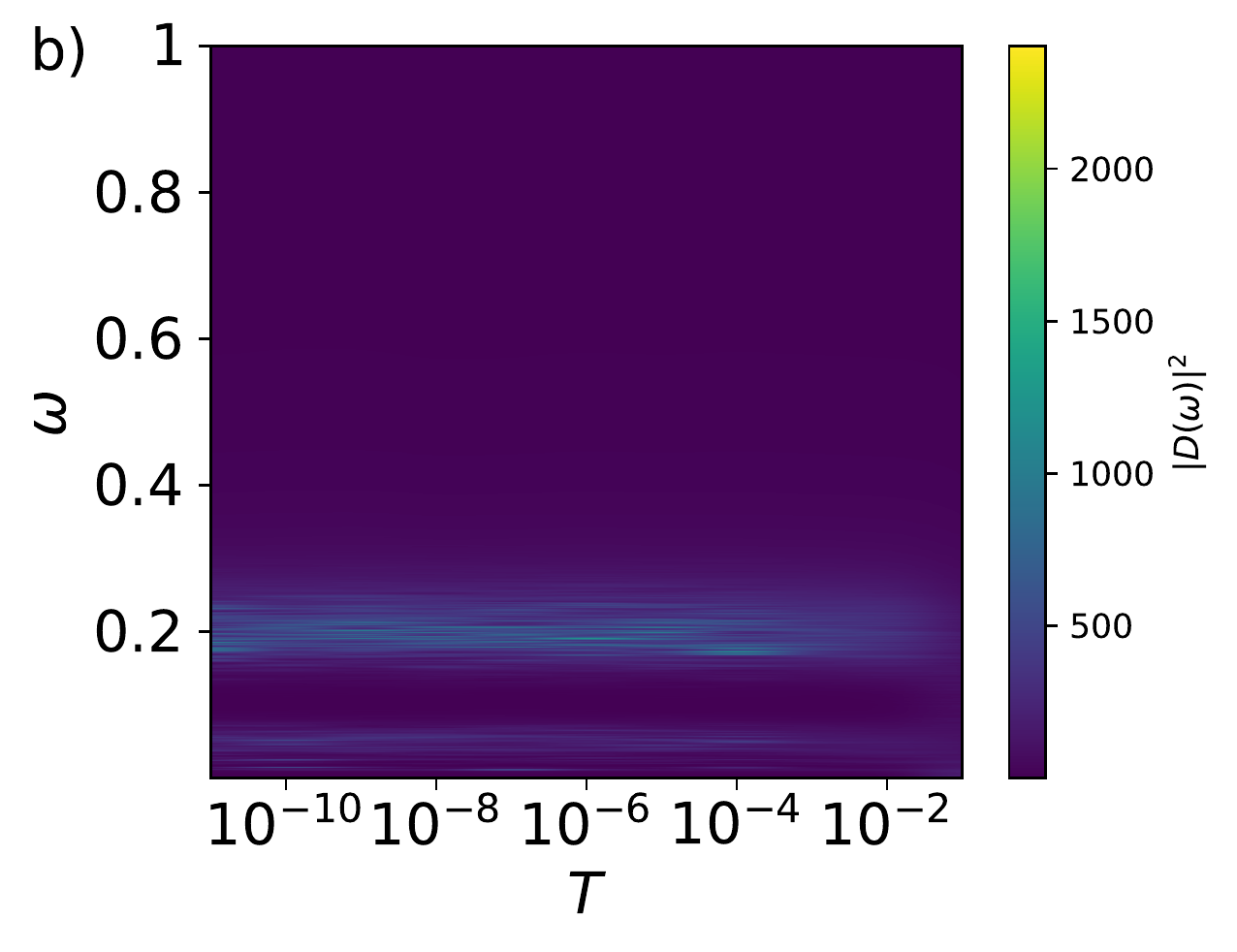}
\includegraphics[height= 4cm]{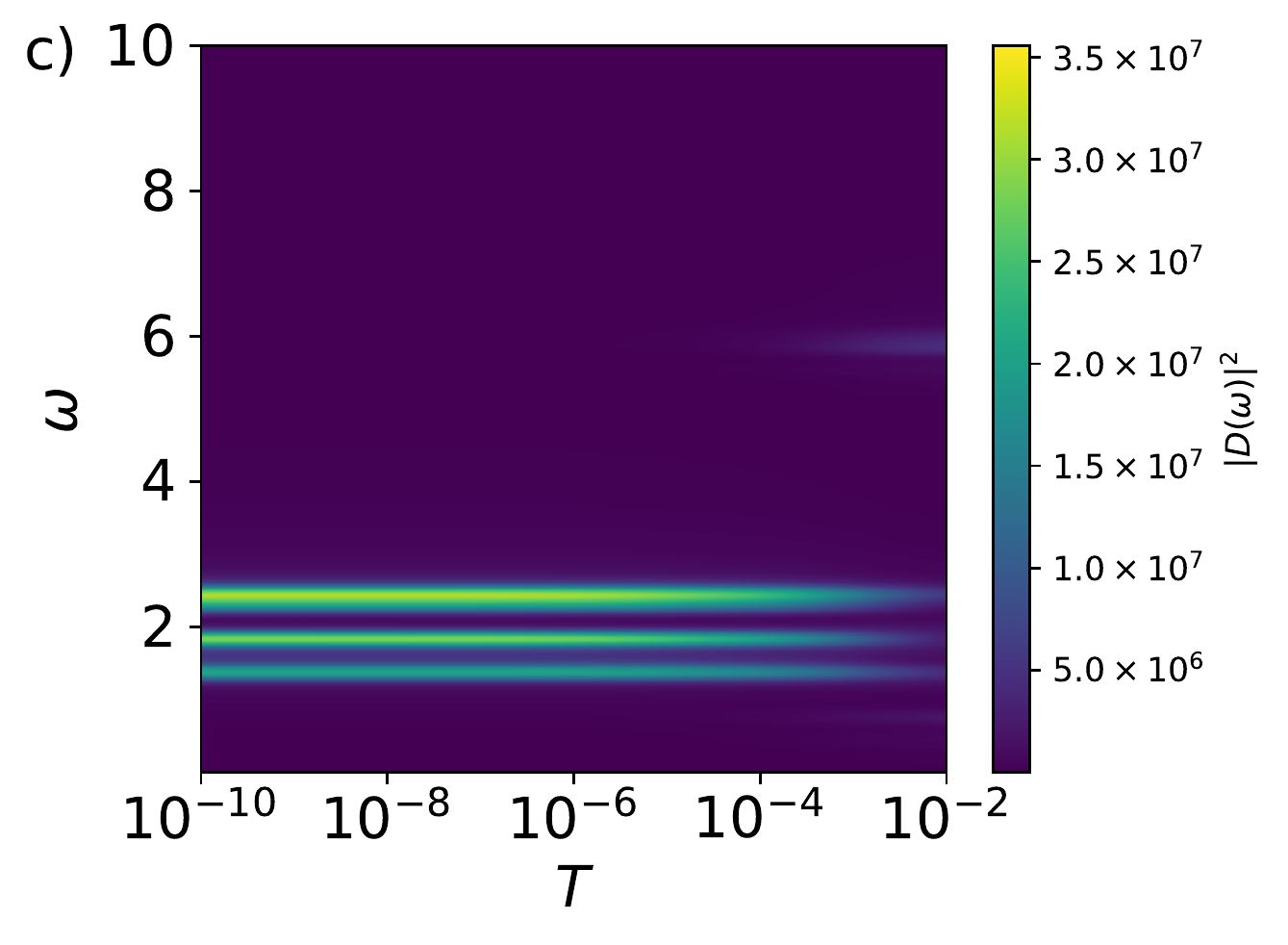}
\includegraphics[height= 4cm]{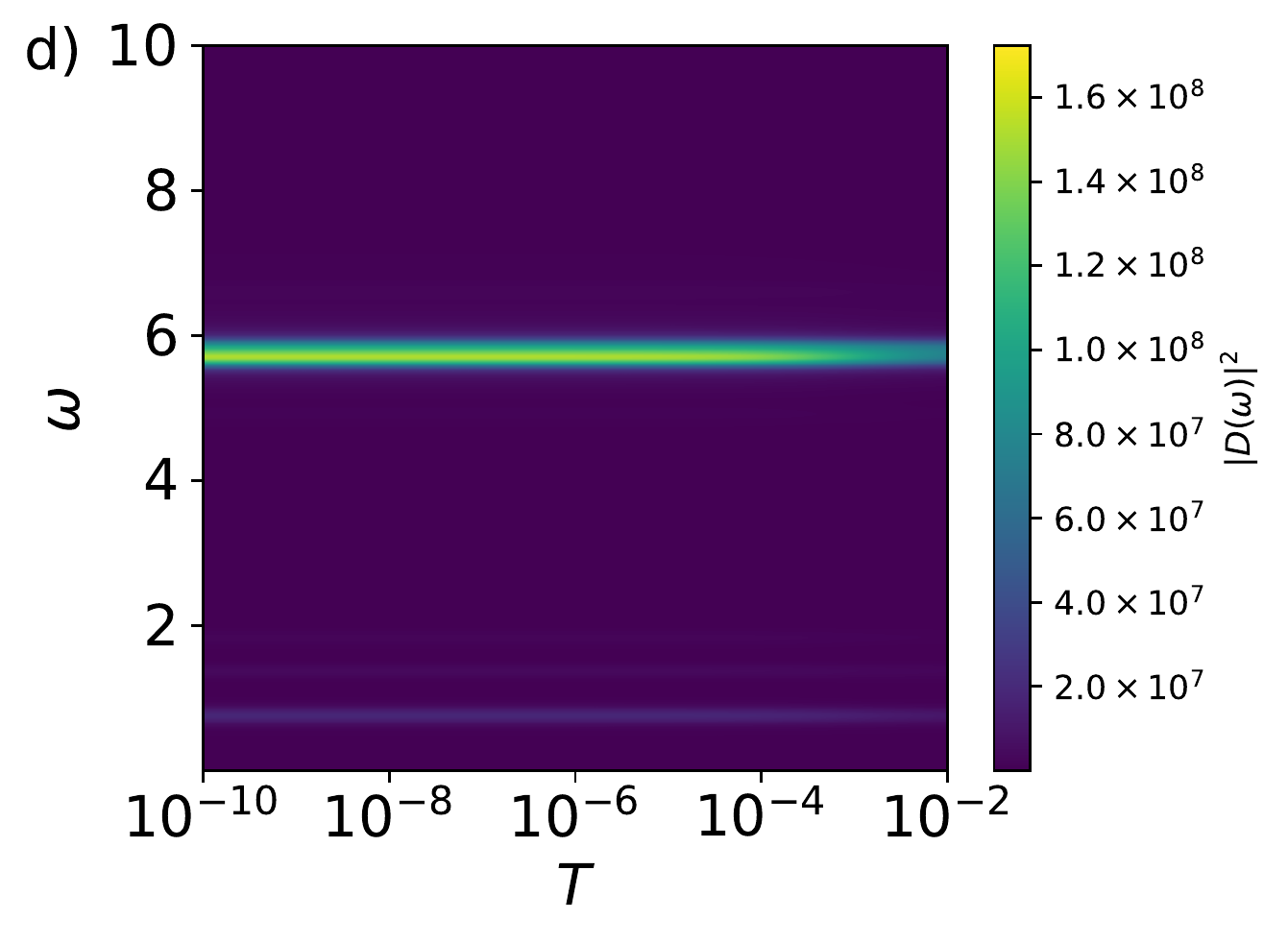}
\caption{Power spectra distribution (Fourier transform of the velocity autocorrelations): for the parabolic potential in the $x$ direction $(a)$ and $y$ direction $(b)$, and the Lennard-Jones potentials in the $x$ direction $(c)$ and $y$ direction $(d)$.}\label{map} %and $y$ directions for the parabolic and the Lennard-Jones potentials.} \label{map}
\end{figure}

In Fig. \ref{fourier}, we can see the corresponding Fourier transform of the panels in Fig. \ref{Velocity}. It is interesting to note that the isotropy properties of the autocorrelation function show up here as well. More specifically, for the Lennard-Jones potential (right panel), the high frequency peak disappears in the $x$ direction for $T < 10^{-4}$. This is illustrated in more detail in Fig. \ref{map}, where $D( \omega )$ is displayed for a range of initial temperatures for both potentials.

Finally, in Fig. \ref{map} we can see the different behavior of $|D(\omega)|^2$ for the parabolic and Lennard-Jones potential in the $x$ and $y$ directions. We can remark the anisotropy that emerges for the Lennard-Jones potential. Notice also that for the Lennard-Jones potential more and stronger frequency peaks appear, compared with the parabolic one.

\section{Conclusions}
Summarizing, we have analysed in this paper the mechanical signal behavior for two granular systems. More concretely, our system is composed by two 
sets of granular particles whose masses are different. Otherwise, the particles have identical properties. Two different potentials have been analysed. They are arranged in an hexagonal lattice
and from the corresponding Hessian matrix we have calculated the inherent eigenfrequencies. In addition, we inspected the behavior of the velocity autocorrelations functions finding important anisotropic behavior in the case of the Lennard-Jones potential for which clearly the transmission is more intense in the $y$ direction (perpendicular to the thermal walls). In addition, the high frequency peak of the velocity autocorrelation Fourier transform disappear in the $x$ direction for the Lennard-Jones potential.

Our results show that the properties of the interaction forces and the initial temperature conditions can significantly change the properties of the mechanical signal transmission in a forced granular system.

%\begin{figure}[ht]
%\includegraphics[width=.85 \columnwidth]{./figs/D.png}
%\caption{Fourier transform of the velocity correlation function.} \label{Fourier}
%\end{figure}

%\section{EIGENVALUES CALCULATION}

%EQUATIONS,

%\section{EIGENFREQUENCIES SPECTRUM}

% \section{PERSPECTIVES}
%\begin{itemize}
%\item[$\bullet$] Study the evolution of the system when different input signals are used.
%\item[$\bullet$] Take into account inelastic collisions \cite{NBC00}.
%\item[$\bullet$] Reproduce the results experimentally.

%\end{itemize}

% \begin{acknowledgment}
%  We acknowledge funding from the Government of Spain through project No. FIS2016-76359-P and from the regional Extremadura Government through projects No. GR18079 \& IB16087, both partially funded by the ERDF.
% \end{acknowledgment}

\bibliography{fonones}
\bibliographystyle{spmpsci}

\end{document}